\documentclass{emulateapj}

\shorttitle{Stellar Dynamics at the Galactic Center}
\shortauthors{Alexander et al.}

\begin{document}

\title{Constraints on the Stellar Mass Function from Stellar Dynamics at the Galactic Center}

\author{Richard~D.~Alexander, Mitchell~C.~Begelman\altaffilmark{1} and Philip~J.~Armitage\altaffilmark{1}}
\affil{JILA, 440 UCB, University of Colorado, Boulder, CO 80309-0440}
\altaffiltext{1}{Department of Astrophysical and Planetary Sciences, University of Colorado, Boulder, CO 80309-0391}
\email{rda@jilau1.colorado.edu}
\email{mitchb@jilau1.colorado.edu}
\email{pja@jilau1.colorado.edu}


\begin{abstract} 
We consider the dynamical evolution of a disk of stars orbiting a central black hole.  In particular, we focus on the effect of the stellar mass function on the evolution of the disk, using both analytic arguments and numerical simulations.  We apply our model to the ring of massive stars at $\simeq0.1$pc from the Galactic Center, assuming that the stars formed in a cold, circular disk, and find that our model requires the presence of a significant population of massive ($>100$M$_{\odot}$) stars in order to explain the the observed eccentricities of 0.2--0.3.  Moreover, in order to limit the damping of the heavier stars' eccentricities, we also require fewer low-mass stars than expected from a Salpeter mass function, giving strong evidence for a significantly ``top-heavy'' mass function in the rings of stars seen near to the Galactic Center.  We also note that the maximum possible eccentricities attainable from circular initial conditions at ages of $<10$Myr are around 0.4--0.5, and suggest that any rings of stars found with higher eccentricities were probably not formed from circular disks.
\end{abstract}



\section{Introduction} 
The relative proximity of the Galactic Center (henceforth GC) provides a unique opportunity for ``close-up'' study of processes that are expected to be crucial in the formation of galaxies and black holes, yet remain essentially unobservable in more distant galaxies.  Recent advances in telescope technology have enabled us to resolve individual stars in the crowded GC environment, and the development of adaptive optics has allowed determination of both velocities and positions of such stars to be made with ever-increasing accuracy \citep[e.g.][]{ghez98,genzel03,ghez05,paumard06}.  These new data have presented several puzzles, notably the presence of a number of B-type stars very close ($\lesssim 0.01$pc) to the GC, and also the detection of one, and possibly two, coherent rings of massive O- \& B-type stars at somewhat larger radii of $\simeq 0.1$pc \citep{genzel03,paumard06}.  These stars are known to be young, and are therefore presumed to be the result of recent star formation at or close to the GC.  The environment at the GC is vastly different from the typical environment of ongoing star formation in the solar neighborhood, with much larger pressures, densities and temperatures, as well as strong tidal forces, so study of the formation of these stars raises a number of interesting issues \citep[see also the recent review by][]{alexander05}.

A popular theory for the origin of these rings of stars is that they formed via fragmentation of accretion disks around the central black hole \citep[e.g.][]{lb03,goodman03,gt04,nayak06}.  The stars are assumed to form on nearly circular orbits, as a result of gravitational instabilities in the disk.  \citet{nayak06} suggests that this will lead to a stellar mass function that is significantly more top-heavy (i.e.~with significantly more massive stars) than that seen elsewhere in the Galaxy, and recent observational studies suggest that this is indeed the case \citep{ns05,paumard06}.  Other suggestions for the formation of these stellar rings exist, however, most notably the infalling star cluster scenario \citep[e.g.][]{gerhard01,mpz03,bh06}.  This scenario can also result in a significantly top-heavy mass function, due to mass segregation and stellar collisions within the cluster \citep*[e.g.][]{bd98,gr05,fgr06}.  

In this paper we consider the dynamical evolution of a ring of stars such as those observed around the GC.  Previous models of the stellar dynamics of such a ring have considered a single stellar mass population \citep{nc05}, or the effect of processes such as resonant relaxation \citep{ha06}.  Here we investigate the effects of the stellar mass function on the evolution of the system, and use observational determinations of the stellar orbital parameters to constrain the mass function of the stellar rings.  We find that the distribution of eccentricities can tell us about both the upper and lower ends of the stellar mass function, and discuss the consequences of this result for theories of star formation in the GC environment.  The structure of the paper is as follows.  In Section \ref{sec:analytic} we present a simple analytic model for the dynamical evolution of a mass-segregated ring of stars orbiting a massive central black hole.  In Section \ref{sec:numerical} we describe numerical simulations of such a system, and compare the results to the predictions of our analytic model.  We then apply our model to the GC system (Section \ref{sec:gc_app}), and derive constraints on the initial conditions by comparing our model to recent observations.  We discuss the consequences of our results, and the limitations of our analysis, in Section \ref{sec:dis}, and summarize our conclusions in Section \ref{sec:conc}.

\section{Analytic model}\label{sec:analytic}
Let us first consider the relaxation of a system of $N_*$ stars of mass $M_*$, orbiting in a disk around a black hole of mass $M_{\mathrm {bh}}$ (where $M_{\mathrm {bh}} \gg N_*M_*$).  The disk (or ring) is centered at radius $R_0$, with a radial width $\Delta R$ and a one-dimensional stellar velocity dispersion $\sigma_*$.  We make the simplifying assumption that the velocity dispersion is isotropic.  This is not strictly valid, but it has been shown that the ratio of the radial and vertical velocity dispersions cannot become larger than 3 without the system becoming unstable \citep*{kmc71,ps77}.  The relaxation time for such a system is given by
\begin{equation}\label{eq:t_rel}
t_{\mathrm {relax}} = \frac{C \sigma_*^3}{G^2 M_* \rho_* \ln \Lambda_*} \, ,
\end{equation}
 \citep[e.g.][]{bt87,pt01} where $\rho_*$ is the stellar density, $\ln \Lambda_*$ is the Coulomb logarithm and $C$ is an order-of-unity constant that depends on the geometry of the system.  (For a spherical system $C\simeq 0.34$, \citealt{bt87}.)  The stellar density is given by
\begin{equation}
\rho_* = \frac{N_* M_*}{2\pi R_0 \Delta R \times 2H} \, ,
\end{equation}
where $H$ is the scale-height of the disk.  We can express $H = \sigma_*/\Omega$ in terms of the stellar velocity dispersion and the orbital angular velocity of the disk, $\Omega$, and so for a Keplerian disk we have
\begin{equation}\label{eq:t_relax}
t_{\mathrm {relax}} = \frac{C_1 R_0 \Delta R \sigma_*^4}{G^2 N_* M_*^2 \ln \Lambda_*} t_{\mathrm {orb}} \, ,
\end{equation}
where $t_{\mathrm {orb}} = 2\pi/\Omega$ is the Keplerian orbital period at $R_0$, and we have absorbed a factor of 2 into the constant such that $C_1=2C$. Consequently, the relaxation of the system is governed by 
\begin{equation}\label{eq:sigma_star}
\frac{d\sigma_*}{dt} = \frac{G^2 N_* M_*^2 \ln \Lambda_*}{C_1 R_0 \Delta R  t_{\mathrm {orb}} \sigma_*^3} \, .
\end{equation}
This form for $d\sigma_*/dt$ allows the velocity dispersion to increase indefinitely, and in reality a cooling term should be included.  However, this cooling arises when stars at the high-velocity tail of the distribution begin to escape from the system, and is only significant once the velocity dispersion becomes comparable to the orbital velocity.  In our application of this model to the GC we are only interested in the behavior of the system at relatively early times, and in this case this simpler treatment remains accurate.  The effects of this simplification are discussed in Section \ref{sec:dis}.

We now extend this analysis to consider two different mass classes of stars, $M_1$ and $M_2$, with $M_1 > M_2$ and velocity dispersions $\sigma_1$ and $\sigma_2$, stellar number $N_1$ and $N_2$, and mean energies per particle $E_1 = 3M_1 \sigma_1^2/2$ and $E_2 = 3 M_2 \sigma_2^2/2$, respectively.  Each mass class is subject to the ``self-relaxation'' described above, but in addition energy can be exchanged between the classes.  This type of analysis has previously been used to study the evolution of a distribution of planetesimals around the sun, and detailed calculations have been made \citep*[e.g.][]{sw88,lissauer93,gls04}\footnote{In the planet formation context, the ``self-relaxation'' is often referred to as ``viscous stirring''.}.  \citet{gls04} point out that the treatment of this problem depends on whether the velocity dispersion of the light stars, $\sigma_2$, is greater than or less than the Hill velocity of the heavy stars, $v_{\mathrm H}$.  The Hill velocity is defined as
\begin{equation}
v_{\mathrm H} = \Omega R_{\mathrm H} \, ,
\end{equation}
where the Hill radius $R_{\mathrm H} = R_0 (M_1/M_{\mathrm {bh}})^{1/3}$.  In the former case, referred to as the ``dispersion-dominated'' regime, the velocity dispersion is a good approximation for the speed of a single star and scattering encounters are well approximated by two-body dynamics.  However, in the latter case, known as the ``shear dominated'' regime, the tidal gravity of the central black hole is important: in this regime the interactions are rather more subtle.

In applying our model to the GC system (see Sections \ref{sec:numerical} \& \ref{sec:gc_app}), we adopt $M_{\mathrm {bh}}=3\times10^6$M$_{\odot}$ and $R=0.1$pc.  Consequently, the Keplerian orbital speed is $v_{\mathrm K} \simeq 360$km s$^{-1}$, and the Hill velocity scales as
\begin{equation}
v_{\mathrm H} = 7.3 \left(\frac{M_1}{25\mathrm M_{\odot}}\right)^{1/3} \mathrm {km\,s}^{-1} \, .
\end{equation}
Thus $v_{\mathrm H} \ll v_{\mathrm K}$, so the Hill velocity corresponds to orbits with very small eccentricities.  In our models, $\sigma_2 > v_{\mathrm H}$ at all but extremely early times, so we work in the ``dispersion-dominated'' regime throughout.  By considering energy conservation in a two-body interaction, we can write the form for the ``exchange'' term as
\begin{equation}\label{eq:ex_term}
N_1 \frac{dE_1}{dt} = - N_2 \frac{dE_2}{dt} = - \frac{6 G^2 N_1 N_2 M_1 M_2 \ln \Lambda_{12}}{C_2 R_0 \Delta R  t_{\mathrm {orb}} \bar{\sigma}_{12}^4} (E_1 - E_2) \, .
\end{equation}
Here $\bar{\sigma}_{12} = (\sigma_1+\sigma_2)/2$ is the mean of the two velocity dispersions (and therefore the mean collision speed), $\ln \Lambda_{12}$ is the appropriate Coulomb logarithm for such a collision, $C_2$ is another order-of-unity constant, and the factor of 6 in the numerator arises because we are considering the energy rather than the velocity dispersion.  We can then use $dE/dt = 3M\sigma d\sigma/dt$ to find expressions for the relaxation of the two velocity dispersions:
\begin{equation}\label{eq:sigma1}
\frac{d\sigma_1}{dt} = \frac{N_1 M_1^2 \ln \Lambda_1}{A_1 t_{\mathrm {orb}} \sigma_1^3} - \frac{N_2 M_1 M_2 \ln \Lambda_{12}}{A_2 t_{\mathrm {orb}}}\frac{\sigma_1}{\bar{\sigma}_{12}^4} \left(1-\frac{E_2}{E_1}\right)
\end{equation}
\begin{equation}
\frac{d\sigma_2}{dt} = \frac{N_2 M_2^2 \ln \Lambda_2}{A_1 t_{\mathrm {orb}} \sigma_2^3} + \frac{N_1 M_1 M_2 \ln \Lambda_{12}}{A_2 t_{\mathrm {orb}}}\frac{\sigma_2}{\bar{\sigma}_{12}^4} \left(\frac{E_1}{E_2}-1\right) \, .
\end{equation}
Here we have rewritten the constant terms for clarity, expressing them as $A_i = C_i R_0 \Delta R/G^2$ for $i=1$,2.  The form of the exchange term can be considered as the product of a relaxation time (containing three powers of ${\bar{\sigma}_{12}}$, as in Equation \ref{eq:t_rel}), a volume scaling term which accounts for the different thicknesses of the two disks ($\sigma_1/{\bar{\sigma}_{12}}$), and a normalized energy difference.  As mentioned above, similar analyses are common in the study of planet formation.  In this context other factors, such as physical collisions, are also significant, but if we neglect these terms we can compare the results to our model.  The form of the solution is the same, and by comparison to the three-dimensional analysis of \citet{sw88} we see that that the constants $C_1$ and $C_2$ are related by the ratio $C_1/C_2 \simeq 3.5$.  We adopt this ratio throughout.

From these equations we can make some qualitative inferences about the evolution of this system.  We see from Equation \ref{eq:t_relax} that the ``self-relaxation'' of the heavy stars will be more rapid than that of the light stars\footnote{Formally, this is true only if $N_1M_1^2>N_2M_2^2$.  However, this holds for any mass function $dN/dM \propto M^{-\Gamma}$ where $\Gamma<3$ (the Salpeter slope is $\Gamma=2.35$), and is therefore true for all cases considered in this paper.}, so we expect the transfer term to boost the velocity dispersion of the light stars, $\sigma_2$, while at the same time damping the velocity dispersion of the heavy stars $\sigma_1$.  However, the rate at which this energy transfer occurs depends on both the velocity dispersions and the distribution of mass in the system.  In Equation \ref{eq:sigma1} the damping term opposes the self-relaxation term when $E_2 < E_1$, and we see that the evolution of the heavy stars can be dominated by the damping term only if $N_2 M_2 \gtrsim N_1 M_1 (\bar{\sigma}_{12}/ \sigma_1)^4$.  Since $\bar{\sigma}_{12}$ is at least of order $\sigma_1$, this means that the damping term can dominate if the total mass in light stars is greater than the total mass in heavy stars.  In the case of the light stars, however, the exchange term acts in the same sense as the self-relaxation term, so we always expect the light stars to relax more rapidly than they would in isolation.  This simple two-component model shows that, in the absence of other factors such as tidal and resonant effects, the stellar mass function is the critical factor in determining the relaxation of a stellar disk.  

In order to make detailed studies of the effect of a real mass function, it is necessary to extend this analysis to at least three mass classes.  We add a third class with $M_3 < M_2$, and consider three exchange terms of the form seen in Equation \ref{eq:ex_term} (1--2, 1--3 and 2--3).  In this case we see that the heaviest stars are damped by both lighter mass classes, while the velocity dispersion of the intermediate stars is boosted by the heaviest stars and damped by the lightest stars, and the lightest stars are boosted by both heavier mass classes.  The distribution of stellar masses is critical to the evolution, but by considering three classes we can now emulate the effect of massive stars, which may only live for some fraction of the lifetime of the system.  We apply this ``three-class'' model to the Galactic Center system in Section \ref{sec:gc_app}.

\section{Numerical Simulations}\label{sec:numerical}
Our analytic model is rather simple, so in order to assess the validity of this approach we have conducted a number of numerical simulations.  We use the $N$-body stellar dynamics code developed by \citet{hm03}, treating both the black hole and the stars as point masses.  This code uses a Hermite integration scheme, which enables us to retain high numerical accuracy.  

Great care must be taken when using an $N$-body code to analyze such a problem, as the energy of the system is overwhelmingly dominated by the central object.  If we assume that the stars have typical separation $H$, then the ratio of the typical energy of a stellar encounter to that of the star--black-hole system is approximately $(M_*/M_{\mathrm {bh}})(R/H)$.  Therefore, for 10M$_{\odot}$ stars orbiting a $10^6$M$_{\odot}$ black hole in a disk with aspect ratio $H/R=0.1$, the energy binding the stars to the black hole is some $10^4$ times larger than the typical energy of stellar encounters.  Consequently, we require very strict limits on the energy errors resulting from the numerical integration if we are to maintain accuracy in our simulations.  This requirement, combined with restrictions on computational time, limit us to modeling systems with relatively few stars, typically 150 or fewer.

\subsection{Single Mass Class}
\begin{figure}
\includegraphics[angle=270,width=\hsize]{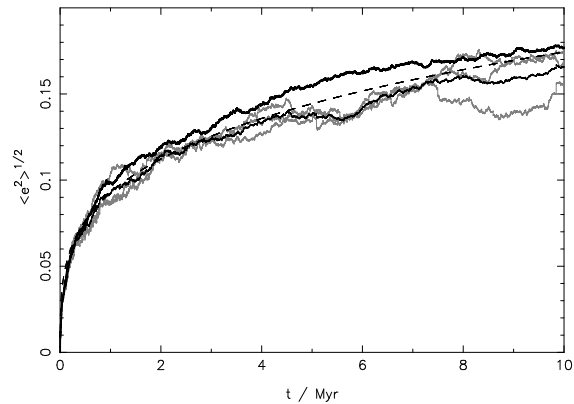}
\caption{Evolution of the rms eccentricity for a system with $M_{\mathrm {bh}}=3\times10^6$M$_{\odot}$, $M_* = 25$M$_{\odot}$ and $N_* = 50$.  The three grey curves show the results of the $N$-body simulations for three different random realizations of the initial conditions, and the black curve is the mean of these three realizations.  The heavy curve shows the mean calculated in the same manner, but with the simulations run with more strict error tolerances (see text).  The dashed curve shows the best-fitting analytic model, with $C_1=2.2$. }\label{fig:ecc_M25_N50}
\end{figure}
We first consider a single mass of stars, in order to determine the value of the numerical constant $C_1$ in Equation \ref{eq:t_relax} (and indeed to test whether a single constant is appropriate).  We consider systems of 50 stars, orbiting around a black hole of mass $3\times10^6$M$_{\odot}$.  We set up the initial conditions as follows.  The stars are distributed in a radial region between $R=0.05$--0.15pc (i.e.~a ring with radius $R_0=0.1$pc and width $\Delta R=0.1$pc), with uniform distributions in both radius and azimuth.  The stars are given a Gaussian distribution in $z$, with scale-height $H=0.05R$.  All stars are given zero-eccentricity Keplerian orbital velocities in the $x$--$y$ plane, with zero velocity in the $z$-direction, and this system is then integrated.

We first consider a stellar mass of $M_*=25$M$_{\odot}$.  We generated three sets of random initial conditions, and integrated each for 6000 orbital periods at $R_0$ ($\simeq 10^7$yr for the parameters specified).  Each model was computed using two different energy error tolerances, in order to check the numerical convergence.  The cumulative fractional energy errors were typically $10^{-9}$ in one case, and $10^{-10}$ in the second case.  Translating the cumulative energy error into a measure of the reliability of the results is not straightforward, but in both cases we consider the estimated energy errors to be sufficiently small.  We computed the orbital elements of each particle twice every orbital period in order to plot the results.

Figure \ref{fig:ecc_M25_N50} shows the evolution of the root-mean-square (henceforth rms) eccentricity in the simulations.  The rms eccentricity is evaluated from the simulations by computing the instantaneous eccentricity of each star directly from the orbital elements.  The discrepancy between the mean values obtained from two sets of simulations (different error tolerances) is around 15\%, and is comparable to the differences between different random realizations of the same simulation.  Furthermore, this discrepancy is also comparable to the typical random fluctuations expected  ($\simeq15$\% for $N_*=50$), so we consider the simulations to be numerically converged.  We see that the rms eccentricity of the stars rises from zero to around 0.15 after 10Myr, and we are able to compare the results of the simulations to the analytic model presented in Section \ref{sec:analytic}.  In a disk of small objects orbiting a massive central body, the rms eccentricity of the small bodies (stars) can be related to their velocity dispersion by
\begin{equation}
e_{\mathrm {rms}} = \sqrt{2} \frac{\sigma_*}{v_{\mathrm K}}
\end{equation}
\citep[e.g.][]{lissauer93} where $v_{\mathrm K} = \sqrt{GM_{\mathrm {bh}}/R_0}$ is the Keplerian orbital speed.  We performed a simple least-squares fit to determine the best-fitting value of the constant $C_1$.  We evaluate the model fit by integrating Equation \ref{eq:sigma_star} numerically.  The Coulomb logarithm is evaluated as the ratio of the maximum to minimum impact parameters.  We assume that the maximum impact parameter is $\Delta R$, and the minimum is $2GM_*/\sigma_*^2$ \citep[e.g.][]{bt87}, and therefore evaluate the Coulomb logarithm as $\Lambda_* = \sigma_*^2 \Delta R / 2 G M_*$.  However, we note that the solutions do not depend strongly on the form adopted for $\Lambda_*$.  For the ``low resolution'' simulations the best-fitting value is $C_1 \simeq 2.55$; the ``high-resolution'' simulations give $C_1 \simeq 1.85$.  However, we note that the velocity dispersion, and therefore the rms eccentricity, depends only on $C_1^{-1/4}$, so this apparently large discrepancy in the scaling constant corresponds only to around a 10\% uncertainty  in the velocity dispersion.  By comparison, the more detailed analysis of \citet{sw88} suggests that $C_1$ lies in the range 1.9--2.8, depending on the degree of anisotropy in the velocity dispersion, so we are satisfied that our simplified analytic form for the ``self-relaxation'' term is accurate to within 10--15\%.

\subsection{Two Mass Classes}\label{sec:two_class}
\begin{figure}
\includegraphics[angle=270,width=\hsize]{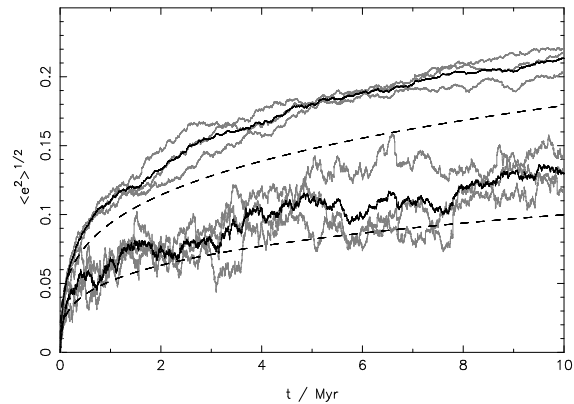}
\caption{Evolution of the rms eccentricity for a system with $M_{\mathrm {bh}}=3\times10^6$M$_{\odot}$, $M_1 = 50$M$_{\odot}$, $N_1 = 10$, $M_2 = 10$M$_{\odot}$ and $N_2 = 100$.  As in Fig.~\ref{fig:ecc_M25_N50}, the grey curves show the results of the $N$-body simulations for three different random realizations of the initial conditions, and the black curve is the mean of these three realizations.  Here the upper set of curves is for the 100 10M$_{\odot}$ stars, and the lower curves are for the 10 50M$_{\odot}$ stars.  The dashed curves again show the best-fitting analytic model, with $C_1=2.2$. }\label{fig:ecc_damp}
\end{figure}
We now consider a system with stars of two different stellar masses, in order to test the accuracy of the exchange term in our analytic model.  We model the evolution of a system with $10\times50$M${\odot}$ stars and $100\times10$M$_{\odot}$ stars.  Thus the total mass in light, 10M$_{\odot}$ stars is double that in heavy, 50M$_{\odot}$ stars, so we expect the eccentricities of the heavy stars to be damped significantly.  For comparison, we also consider a system with $10\times50$M${\odot}$ only.  Once again we evaluated three random realizations of the initial conditions for each; the typical cumulative fractional energy errors were $\sim 10^{-9}$.  However, with many more stars it was not practical to run a full convergence test in this case.  Runs with more stringent error constraints, which were run for much shorter times, suggest that these simulations have relaxed somewhat too rapidly, and that the accuracy of these simulations is around $20$\%.  However, we note that this comparison only considers the early evolution, where the velocity dispersion is rising very steeply.

The results of the two-class simulations are shown in Fig.~\ref{fig:ecc_damp}.  The analytic model provides an excellent fit to the comparison ($10\times50$M${\odot}$ only) simulations and, as expected, the eccentricities of the heavy stars in the two-class model are significantly damped.  The analytic curves are slightly below the simulated data, but we suggest that this is a result of slightly insufficient numerical accuracy in the simulations.  Even with this caveat, the agreement is good both in terms of the relative eccentricities of the two mass classes and the absolute numerical scaling.  Consequently, we are confident that our simple analytic model is accurate to within $\simeq 15$\%.

\section{Application to the Galactic Center}\label{sec:gc_app}
We now apply our analysis to the GC system.  One of the more surprising results of recent years was the detection of large numbers of young stars very close to the GC, and these stars have now been well-studied observationally.  A population of massive O and B stars (sometimes referred to in the literature as the GC ``He\,{\sc i} stars'') is known to exist in one, and possibly two, coherent ring-like structures at a distance of around 0.1pc from the GC \citep[e.g.][]{genzel03,ghez05,paumard06}.  The young age \citep[$\sim 6$Myr,][]{paumard06} of these stars suggests that they formed at or very close to their current location, but the environment at the GC poses a significant challenge to conventional theories of star formation.  Moreover, the detection of similar rings of stars in the center of M31 \citep{bender05} suggests that such systems may in fact be common, so their evolution warrants further study.

One important goal of any theory of star formation is to predict the form of the initial stellar mass function [henceforth (I)MF].  Observational limitations (primarily source confusion) limit the study of the lower-end of the mass function at the GC through direct observations, but several indirect approaches have been taken to try to determine its form.  \citet{paumard06} infer the form of the MF from their observed $K$-band luminosity function, and conclude that the slope of the MF $\Gamma$ is in the range $\Gamma = 0.85$--1.35 (i.e.~$dN/dM \propto M^{-\Gamma}$; in these units the Salpeter slope is $\Gamma=2.35$).  \citet{ns05} argue that the integrated X-ray luminosity of the GC region observed by {\it Chandra} sets a limit on the total mass in young, low-mass T Tauri stars, of $\lesssim10^4$M$_{\odot}$.  We note, however, that young stars of earlier spectral type emit a much smaller fraction of their luminosity in X-rays than their T Tauri counterparts \citep[e.g.][]{stelzer06}, so formally this limit only applies to stars less massive than approximately 5M$_{\odot}$.  \citet{ndcg06} consider the mutual interaction of the two stellar rings observed by \citet{genzel03} and \citet{paumard06}, and suggest that dynamical interactions between the two rings would destroy the observed structure unless the total mass of the rings was $\lesssim 10^4$M$_{\odot}$.  Finally, \citet{nc05} argue that $N$-body interactions within a stellar ring (without considering mass segregation) set a somewhat weaker upper limit on the total stellar mass, $\lesssim 3\times10^5$M$_{\odot}$.  These studies, combined with the knowledge that $\sim 3000$M$_{\odot}$ is present in more massive early-type stars \citep{genzel03,paumard06}, are strongly suggestive of a significantly top-heavy MF in the GC system, with many fewer low-mass ($\lesssim 5$M$_{\odot}$) stars than would be expected from a standard Salpeter MF.  \citet{nayak06} suggested that such a top-heavy initial MF will arise naturally if stars are formed by the fragmentation of an accretion disk around the central BH, and models of the ``infalling cluster'' scenario predict a similarly top-heavy MF \citep{gr05,fgr06}.  Here we investigate the formation of the GC system further by considering the effect of the MF on the dynamical evolution of a stellar ring.

\subsection{Basic Model}\label{sec:basic}
\begin{figure}
\includegraphics[angle=270,width=\hsize]{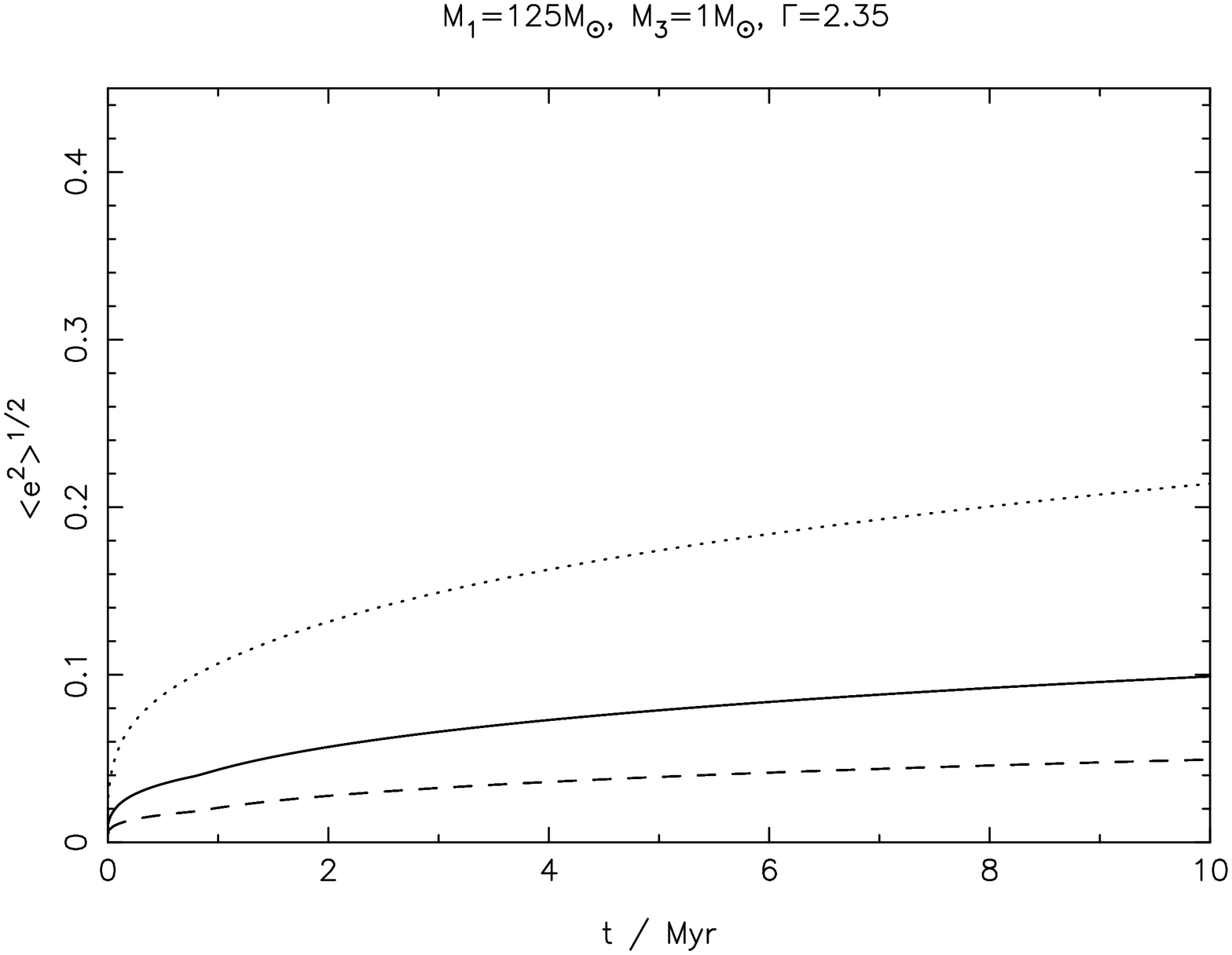}

\includegraphics[angle=270,width=\hsize]{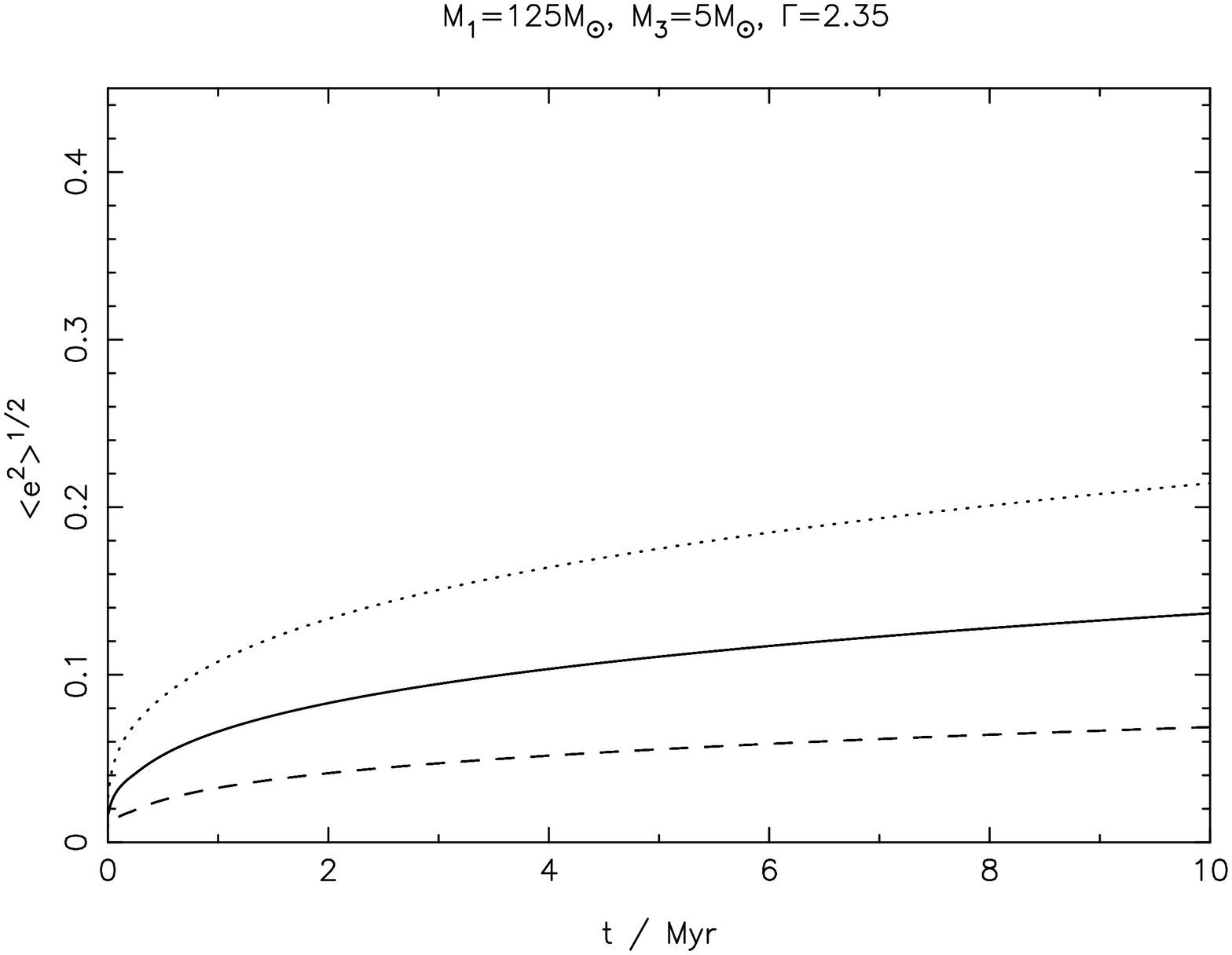}

\includegraphics[angle=270,width=\hsize]{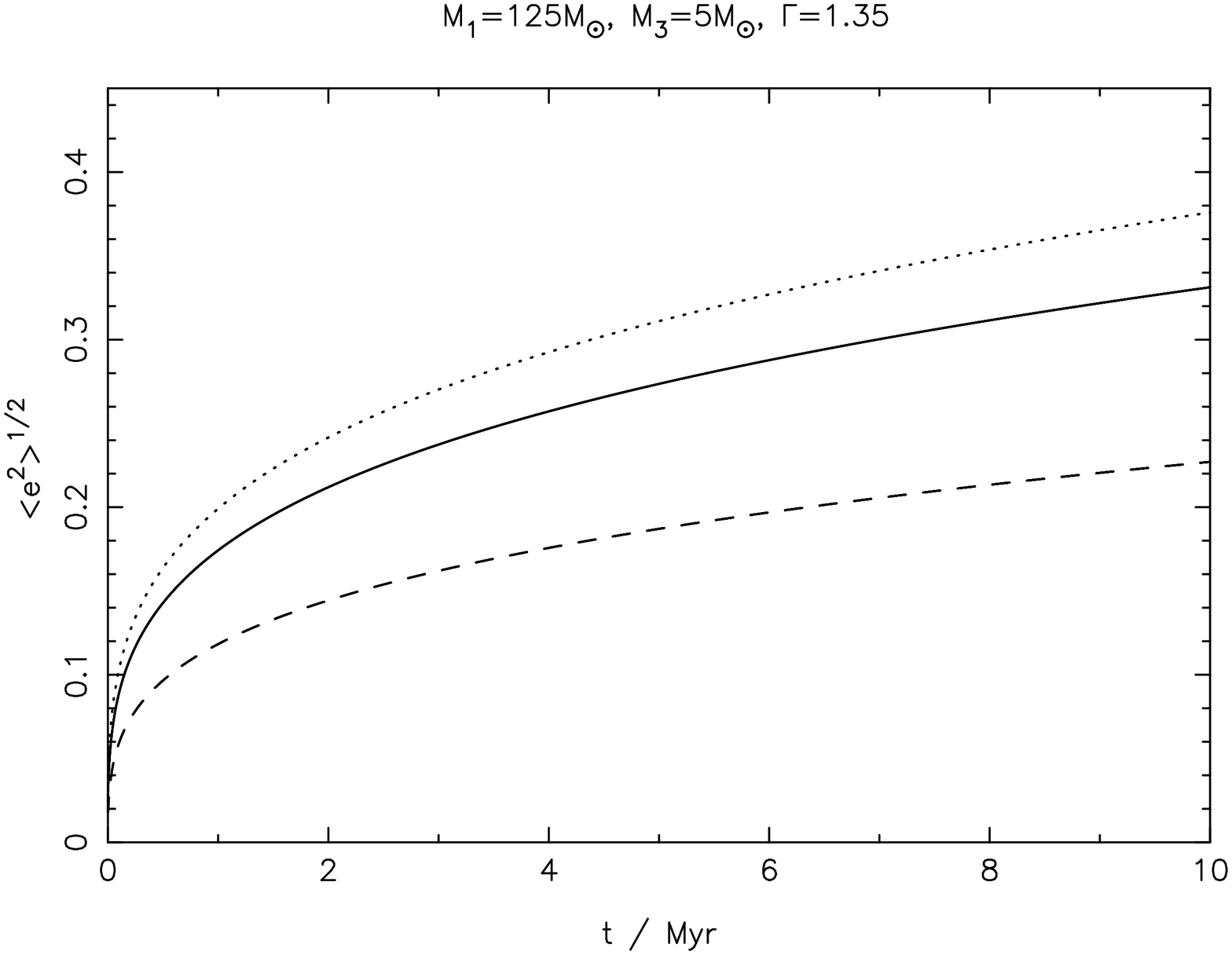}
\caption{Evolution of rms eccentricity in our ``three class'' model.  In all three panels the dashed, solid and dotted lines represent the three mass classes from heaviest to lightest respectively, and in each case $M_2 = 25$M$_{\odot}$ and $N_2=50$.  The upper panel shows the evolution for a Salpeter mass function with $M_1=125$M$_{\odot}$ and $M_3=1$M$_{\odot}$: in this case we see that the light stars damp the eccentricities of the heavier stars significantly.  The middle panel shows the same mass function slope $\Gamma=2.35$, but this time with a low-mass cutoff at $M_3=5$M$_{\odot}$: we see that the damping effect is lessened in this case.  The lower panel shows a top-heavy mass function, with $\Gamma=1.35$: in this case the presence of many more massive stars results in much larger eccentricities.}\label{fig:threeclass1}
\end{figure}
In order to study the effect of the stellar MF on the dynamical evolution of the GC system we use the ``three-class'' model presented in Section \ref{sec:analytic}.  \citet{paumard06} identify 53 OB and Wolf-Rayet (WR) stars in the clockwise stellar ring, and we use this observation to fix the central mass class in our model.  The OB and WR stars have masses in the range $\simeq 20$--30M$_{\odot}$, so we adopt $M_2 = 25$M$_{\odot}$ and $N_2 = 50$.  We model the effect of a varying mass function by allowing the masses $M_1$ and $M_3$, and the slope of the mass function, $\Gamma$, to be free parameters in our model.   We choose $N_1$ and $N_3$ according to the number of stars prescribed by the MF at exactly $M_1$ and $M_3$, and thus the choice of the MF slope fixes $N_1$ and $N_3$.
\citet[][see also \citealt{bel06}]{paumard06} find the rms eccentricity of the stars in the clockwise ring to be 0.2--0.3.  However, they find that the less well-defined counter-clockwise ring (containing around 20 stars) has significantly larger eccentricities, with an rms value of around 0.6--0.7.  However, we clearly see from Fig.~\ref{fig:ecc_M25_N50} that a ring of $50\times25$M$_{\odot}$ stars will reach a peak rms eccentricity of only 0.15 in 4--8Myr.  Assuming that the stellar orbits were originally circular \citep[as expected from formation in an accretion disk:][]{nayak06}, we therefore require the presence of more massive stars in order to further excite the eccentricity of these stars to the level seen at the GC.

We choose three initial parametrizations of our model. Firstly, we choose a simple Salpeter MF ($\Gamma=2.35$), with a maximum mass $M_1=125$M$_{\odot}$ and a minimum mass $M_3=1$M$_{\odot}$.  We then consider a Salpeter slope with a ``low-mass cutoff'', adopting $M_3=5$M$_{\odot}$.  (As the total mass of a Salpeter MF diverges to low mass, this ``cutoff'' results in a lower total stellar mass.)  Thirdly, we consider a significantly flatter MF, similar to that found by \citet{paumard06} ($\Gamma=1.35$), while maintaining $M_1=125$M$_{\odot}$ and $M_3=5$M$_{\odot}$

\begin{figure}
\includegraphics[angle=270,width=\hsize]{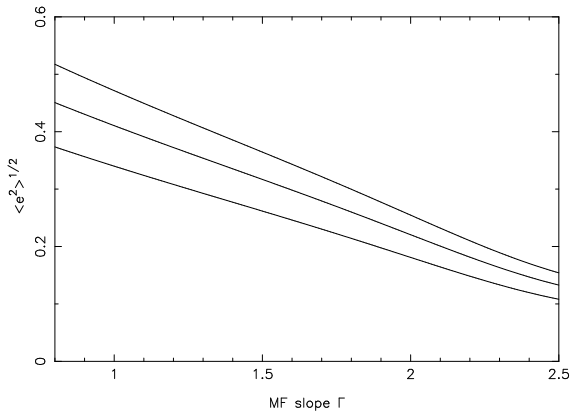}
\caption{``End state'' rms eccentricities of the $50\times25$M$_{\odot}$ stars as a function of the MF index $\Gamma$.  $M_1=125$M$_{\odot}$ and $M_3=5$M$_{\odot}$ are fixed, and the three lines show the rms eccentricity of the 25M$_{\odot}$ stars at 3.0, 6.0 and 10.0Myr (bottom to top, respectively).  We see that a significantly top-heavy MF is required in order to excite the eccentricity of the 25M$_{\odot}$ stars to the level seen in the clockwise system \citep{paumard06}, and also that the maximum rms eccentricity attainable is around 0.4--0.5.}\label{fig:gamma_ecc}
\end{figure}

The results of these models are shown in Fig.~\ref{fig:threeclass1}.  We find that a standard Salpeter MF results in significant damping of the eccentricities of the more massive stars.  This is not altogether surprising, as the Salpeter mass-function diverges to low mass, but we see clearly from Fig.~\ref{fig:threeclass1} that the rms eccentricity of the 25M$_{\odot}$ stars is damped to $\lesssim0.1$ at the age of the GC system.  Even allowing for the uncertainties both in our model and in the observed data, this is significantly lower than the observed eccentricities, and essentially rules out a standard IMF for the GC rings if the stars were initially on circular orbits.  When the low end of the Salpeter MF is truncated at a higher mass the rms eccentricity of the 25M$_{\odot}$ stars suffers noticeably less damping, but (as noted above) the rms eccentricity is still somewhat lower than observed, as there are not enough massive stars present to excite the eccentricities further.  Only when a significantly flatter MF is adopted, resulting in a much larger number of 125M$_{\odot}$ stars, do the 25M$_{\odot}$ stars reach eccentricities of $\simeq0.2$ in 5Myr, as demanded by observations.  We also note that these low to moderate eccentricities mean that few, if any, stars will be ejected from the ring over 10Myr timescales, even at low stellar masses.

Fig.~\ref{fig:gamma_ecc} shows the effect of varying the mass function slope on the rms eccentricity of the 25M$_{\odot}$ stars, with $M_1=125$M$_{\odot}$ and $M_3=5$M$_{\odot}$.  We clearly see that a Salpeter MF lacks sufficient massive stars to excite the rms eccentricity above 0.2.  However, even extremely top-heavy mass functions fail to excite the eccentricities to values much greater than $\sim0.5$ at the age of the GC system, which suggests that the more eccentric counter-clockwise system \citep{paumard06} may pose a problem for our model.  We return to this issue in Section \ref{sec:implications}.

\subsection{Mass-loss from massive stars}
\begin{figure}
\includegraphics[angle=270,width=\hsize]{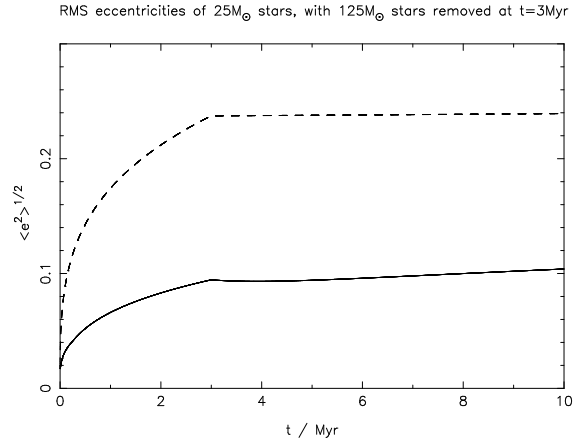}

\includegraphics[angle=270,width=\hsize]{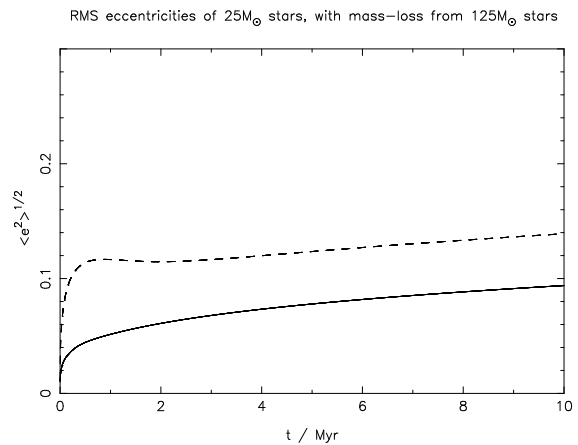}
\caption{Evolution of rms eccentricity when mass-loss is taken into account.  For clarity we plot only the eccentricities of the 25M$_{\odot}$ (``class 2'') stars.  The upper panel shows the case where the heaviest (125M$_{\odot}$) stars are simply removed at $t=3$Myr.  The solid line shows the case of a Salpeter MF, truncated at $M_3=5$M$_{\odot}$; the dashed line shows the case of a top-heavy mass function ($\Gamma=1.35$), also with the same low-mass cutoff.  The lower panel is the same, but in this case the heaviest stars were subject to progressive mass-loss as specified in Equation \ref{eq:massloss}.  In both cases we see that a top-heavy IMF is required in order to produce rms eccentricities (in the 25M$_{\odot}$ stars) larger than around 0.15.}\label{fig:massloss}
\end{figure}
This simple treatment of the GC system is unlikely to be valid, however, as stars born at the upper end of the IMF are subject to significant mass-loss during their lifetimes.  Moreover, it is unlikely that stars of greater than $\sim 50$M$_{\odot}$ will live for longer than 2--3Myr \citep[e.g.][]{schaller92}.  The effect of a supernova blast wave on the dynamics of the stellar rings is negligible \citep*[e.g.][]{wlm75}, but the removal of the massive stars from the system can have a significant effect, as they are no longer able to excite the velocity dispersion of lower mass stars.  We use two crude approximations to account for mass-loss and/or stellar death in our simple model.

Our first approach is simply to assume that the most massive stars are not subject to any mass-loss, but are removed from the system at the end of their lifetimes.  We adopt a lifetime of 3Myr for these stars, and model this by setting $N_1=0$ from $t=3$Myr onwards.  Our second approach is to include mass-loss from the most massive stars.  This is done simply by making the stellar mass $M_1$ a time-dependent function $M_1(t)$.  The details of mass loss from very massive stars are not well understood, but the qualitative effect of mass-loss on the dynamical evolution of the system is not strongly dependent on the mass-loss rates adopted.  For simplicity we adopt the mean mass-loss rate for WR stars derived by \citet{nl00}:
\begin{equation}\label{eq:massloss}
\log_{10}\dot{M}_1 = -5.73 + 0.88\log_{10}M_1 \, .
\end{equation}
We use this form for the mass-loss rate $\dot{M}_1$ for $M_1 > 25$M$_{\odot}$, and assume that the stars' mass remains constant once it drops to 25M$_{\odot}$.  This results in a rather rapid decline in the mass of the heaviest stars, with the mass $M_1(t)$ falling from $125$M$_{\odot}$ to 25M$_{\odot}$ in 1.25Myr.  We note, however, that this form was derived from observations of WR stars, the most massive of which were around 50M$_{\odot}$, and so is of uncertain validity at very large stellar mass.  

The results are shown in Fig.~\ref{fig:massloss}.  We see that the consequence of including mass-loss is, as expected, that the eccentricities of the observable stars are not excited as highly as in the case when no mass-loss was included.  The exact eccentricities achieved are rather sensitive to some rather poorly-defined parameters in the model, namely the mass-loss rates and also the upper cutoff of the MF.  However, in broad terms we see that the typical eccentricities of the observable stars are expected to be in the range $\simeq0.1$--0.3, depending on the details of the upper end of the MF.

\subsection{Implications for the Galactic Center system}\label{sec:implications}
Taken together, these results have important consequences for the GC system. As noted above,  \citet{paumard06} find typical eccentricities in the clockwise ring of around 0.2--0.3, and in the counter-clockwise ring of $\simeq0.7$.  Our analysis has shown that, in order for the observed stars to reach even the moderate eccentricities seen in the clockwise ring, there must be a significant population of much more massive ($>100$M$_{\odot}$) stars present in numbers greatly exceeding those expected from a Salpeter MF.  We also find that the total mass in low-mass ($\lesssim 5$M$_{\odot}$) stars must be less than the mass in OB stars in order to avoid significant damping of the eccentricities of the OB stars.  Thus we conclude that the clockwise system has a rather top-heavy mass function, in agreement with the previous observations of \citet{ns05} and \citet{paumard06}, and find that the observed eccentricities are in agreement with those expected from star formation in a Keplerian disk \citep[e.g.][]{gt04,nayak06}.

As noted in Section \ref{sec:basic}, however, the much larger eccentricities in the counter-clockwise system \citep{paumard06,bel06} pose a problem for our model.  Even in a ``best-case'' model, with no mass-loss from the heaviest stars, and a very top-heavy MF ($\Gamma=0.85$) extending to very high stellar mass (175M$_{\odot}$), the observable 25M$_{\odot}$ stars only reach an rms eccentricity of 0.5 in 10Myr, and more realistic models suggest that rms eccentricities greater than 0.4 are unlikely (see Fig.~\ref{fig:gamma_ecc}).  We therefore conclude that the counter-clockwise system was probably not formed from an initially circular disk, and suggest that some other mechanism must be responsible.

As noted above, our model assumes initially circular stellar orbits, but it may be possible to produce larger eccentricities if the initial configuration of the system has significant eccentricity.  Such a configuration is also possible in the infalling cluster scenario \citep{gerhard01,mpz03}.  Within the disk fragmentation scenario it may be possible to generate eccentric initial conditions if the disk is itself eccentric when it fragments.  If the disk is formed by some individual ``accretion event'', such as the capture of a molecular cloud \citep[as suggested by ][]{nayak06}, then an eccentric stellar disk can form if the fragmentation timescale is shorter than the circularization timescale of the disk.  Models of eccentric accretion disks have shown that a disk can typically remain eccentric for many orbital times \citep{sc92,ogilvie01}, so rapid fragmentation of such a disk may indeed provide a mechanism for generating the large observed stellar eccentricities.  We note also that eccentric initial conditions may retain an observable signature, in the form of a significantly anisotropic velocity dispersion.  Observations of the stellar disk(s) at the center of M31 suggest significant, coherent, eccentricities, and models of eccentric stellar disks have been shown to fit the observed data well \citep[e.g.][]{tremaine95,pt03,bender05}.  We note that the central black hole in M31 is approximately 100 times more massive than that in the Galaxy, so an eccentric stellar disk will retain a preferred eccentricity vector for a much longer time in M31 than at the GC.  However, detailed study of this problem, and associated issues to do with the evolution of initially coherent eccentric stellar disks, is beyond the scope of this investigation.

Our model predicts that few, if any, stars should be scattered inwards from the rings at 0.1pc, and we therefore make no attempt to explain the origin of the so-called ``S-stars'' \citep[e.g.][]{genzel03,ghez05}, which orbit much closer to the GC (at a radius of $\simeq 0.01$pc).  It has been suggested that these stars may have been scattered inwards from the rings by an intermediate-mass ($10^3$--$10^4$M$_{\odot}$) black hole \citep{hansen03}, but our model does not make any prediction in this regard.  We note, however, that such an interaction would provide an additional means of exciting eccentricity in the stellar rings, and therefore may provide another means of generating the large eccentricities observed in the counter-clockwise ring.

\section{Limitations}\label{sec:dis}
There are several obvious limitations to our analysis.  Our model assumes that the velocity dispersion is isotropic, and while this assumption is not unreasonable it is not strictly valid.  Similar analyses applied to planetesimals \citep[e.g.][]{sw88} suggest that in equipartition the radial velocity dispersion is roughly double the vertical dispersion.  Our simulations are complicated by the fact that our initial conditions have zero velocity dispersion in the radial direction (Keplerian orbits), but non-zero dispersion in the vertical 
direction due to the finite thickness of the ring.  The system does not reach equipartition over the timescales considered, but the general trend seems to be in broad agreement with previous analyses.  This suggests that our relation between the velocity dispersion and the disk thickness is not exact, but we note that the manner in which we fit the scaling constant $C_1$ enables us to account for this.  Furthermore, we find good agreement between the model and the simulations, so we do not consider anisotropy in the system to be a significant problem.

As mentioned in Section \ref{sec:analytic}, our analytic model does not allow for cooling, and consequently does not permit an equilibrium solution.  As long as the velocity dispersion remains small relative to the orbital speed (i.e.~the rms eccentricity remains small) this approximation is valid, and this is supported by the favorable comparison between the analytic model and the $N$-body simulations.  This approach may result in an over-estimate of $\sigma$ (and therefore the eccentricity) if $\sigma$ becomes large.  However, the largest eccentricities attained by any of our models are $\simeq 0.5$, so we do not consider this to be a significant problem.  
We also neglect cooling via physical stellar collisions, but note that at the stellar densities considered here this is unlikely to be significant.
Similarly, although we consider only the ``dispersion dominated'' regime for our two- and three-class models, we note that our initial conditions (with circular orbits) lie in the``shear-dominated'' regime.  However, the relaxation of the system to the dispersion-dominated regime is very rapid, typically occurring within $\sim 100$ orbits, so we are satisfied that this simplification does not affect the results significantly.

As noted in Section \ref{sec:two_class}, the convergence of our ``two-class'' numerical simulations is rather marginal, with tests indicating that the relaxation of the system is somewhat too fast.  Unfortunately it is not practical to run more stringent simulations with this type (Hermite) of numerical algorithm, and is it not clear if more efficient algorithms (such as tree-codes) will provide sufficient accuracy to study this problem.  Our tests suggest that the accuracy of our simulations is likely no better than $\pm 15$\%, but we  are satisfied that this level of uncertainly does not affect the qualitative results of the model.  We also note, as above, that our treatment of mass-loss from massive stars is somewhat arbitrary.  However, the two models chosen span a significant fraction of the available parameter space, and more extreme parametrizations do not alter our conclusions significantly.

As noted in Section \ref{sec:analytic}, our model neglects other dynamical effects such as tidal and resonant effects, and also treats the black hole as a point mass with a Newtonian potential.  At the large radius considered here (0.1pc) the timescale for relativistic precession of orbits is much longer than the age of the system, so we can safely neglect relativistic effects.  (The expected precession rate is a few degrees in 10Myr: \citealt*{wmg05}.) Moreover, the non-Keplerian component of the potential due to either remnant black holes \citep{mg00} or other stellar populations at the GC \citep[e.g.][]{genzel03,ghez05} is not expected to be significant. The effect of resonances on the GC system has recently been studied by \citet{ha06}, who find that resonant relaxation can in fact dominate over the type of uncorrelated two-body interactions considered here.  They note, however, that the stellar disks observed at the GC are sufficiently young as to be unaffected by resonant effects.  The fact that our model, which neglects resonant effects, provides a good fit to our numerical simulations supports this conclusion.  It may well be that resonant effects will dominate the future evolution of the stellar rings at the GC, but in the early stages of evolution considered here they do not.


\section{Summary}\label{sec:conc}
In this paper we have considered the dynamical evolution of rings of stars around a massive black hole.  Through analytic arguments and numerical simulations we have constructed a model for the evolution of a disk of stars of different masses, and shown that the stellar mass function is the dominant factor in determining the evolution of such a system.  We have then applied our analysis to rings of stars observed to orbit the  Galactic Center system.  We find, in agreement with previous studies, that the total mass in low-mass ($\lesssim 5$M$_{\odot}$) stars must be significantly lower than expected from a Salpeter mass function, and also find that a significant population of massive ($>100$M$_{\odot}$) stars must have been present in order to produce eccentricities in the range 0.2--0.3, as observed by \citet{paumard06}.  However, we find that dynamical relaxation alone is unlikely to produce rms eccentricities larger than $\simeq 0.4$ in the GC system.  Consequently we conclude that rings with larger eccentricities, such as the counter-clockwise system observed by \citet{paumard06}, are unlikely to have originated in a circular disk, and suggest that some other dynamical process must be responsible for such systems.  Alternatively, we suggest that star formation by fragmentation of an eccentric accretion disk could produce the observed eccentricities.


\acknowledgments
We acknowledge useful discussions with Elena Rossi during the initial stages of this work, and we thank both the editor, Frederic Rasio, and an anonymous referee for useful comments.  This work was supported by NASA under grants NAG5-13207, NNG04GL01G and NNG05GI92G from the Origins of Solar Systems, Astrophysics Theory, and Beyond Einstein Foundation Science Programs, and by the NSF under grants AST--0307502 and AST--0407040.  


\end{document}